\documentclass{article}
\usepackage{epsfig}

\title{Membrane bound protein diffusion viewed by fluorescence 
recovery after bleaching experiments : models analysis.}
\date {\small 2002, july 24th}
\author
{C. Favard \thanks{e-mail : favard@ipmc.cnrs.fr},
 N. Olivi-Tran, J.-L. Meunier  \\ \small CNRS - Institut de Pharmacologie Moleculaire et Cellulaire, \\ \small 660, route des lucioles,
06560  Valbonne Sophia-Antipolis\\ \small
CNRS - Institut Non Lineaire de Nice, \\ \small1361, route des lucioles, 
06560 Valbonne Sophia-Antipolis \\}

\begin{document}

\maketitle

\begin{abstract}

Diffusion processes in biological membranes are of interest to 
understand the macromolecular organisation and function of several molecules. 
Fluorescence Recovery After Photobleaching (FRAP) has been widely used as a 
method to analyse this processes using classical Brownian diffusion model. In 
the first part of this work, the analytical expression of the fluorescence 
recovery as a function of time has been established for anomalous diffusion 
due to long waiting times. Then, experimental fluorescence recoveries recorded 
in living cells on a membrane-bound protein have been analysed using three 
different models : normal Brownian diffusion, Brownian diffusion with an 
immobile fraction and anomalous diffusion due to long waiting times.

\end{abstract}

\section{Introduction}

Early models of the plasma membrane, notably the fluid mosaic model \cite{b.a} 
postulated that transmembrane proteins were freely diffusing in a sea of 
lipids. During this last decade, it has become apparent that cell surface 
membranes are far from being homogeneous mixture of their lipid and protein 
components. They are compartimented into domains whose composition, 
physical properties and function are different. Numerous studies on 
transmembrane proteins by means of single particle tracking (SPT) or 
fluorescence recovery after photobleaching (FRAP) has shown the existence of 
micrometer and submicrometer size domains on both model membrane and living 
cells \cite {b.b,b.c,b.d,b.e}. Kusumi {\it et al. \it} recently formulated the 
"matrix" or "skeleton fence" model for hindered protein movements in which 
transmembrane proteins are coralled by a fence of cytoskeleton just beneath 
the membrane \cite {b.d}. 

FRAP experiments have been used for determination of long-range molecular 
diffusion of proteins and lipids on both model system and cells for more than 
30 years \cite {b.f,b.g}. Briefly, fluorescently labelled molecules localized 
within a predefined area are irreversibly photo-destructed by a short and 
intense laser pulse. The recovery of the fluorescence in this area is then 
measured against time. Since no reversible photoreaction occurs, recovery of 
the fluorescence in the photobleached area is due to diffusion. FRAP data are 
generally interpreted by assuming classical Brownian diffusion. Two parameters 
can then be obtained : D, the lateral diffusion coefficient and M, the mobile 
fraction of the diffusing molecule. When the radius of the photobleached area 
is small compared to the diffusion area, M must be equal to 1 for freely diffusing species.
In fact, most of the data reported so far in biological membranes for transmembrane 
proteins shows a value of $M < 1$. This lack in total fluorescence recovery 
can be interpreted as a restriction to free-diffusion behaviour. Parameters 
obtained have then to be re-evaluated to recognize the effect of time-dependent 
interactions in a field of random energy barriers.

Membrane bound proteins should also be submitted to several interactions
with their surrounding that should account 
for an anomalous subdiffusion behaviour.  Sources of restriction to free 
diffusion may include lipid domains trapping, binding to immobile proteins 
and/or obstruction by cytoskeletal elements.
Therefore, in this letter, diffusion of an intracellular 
membrane-bound protein domain (pleckstrin homology domain) has been analysed inside living cells 
by FRAP experiments. Previous structural studies have shown that these proteins 
are linked to the polar head of specific lipids by means of 
electrostatic interactions \cite{b.h}. Furthermore, the protein used in this study
(PH domain of Exchange Factor for ARF6) appears to have a 
functional requirement to be associated to the plasma membrane within cells 
\cite{b.i}. After an analytical determination of the fluorescence recovery function based on  
an anomalous subdiffusion model, experimental recoveries obtained in living 
cells were analysed using random diffusion with or 
without an immobile fraction and compared to the analyse using time-dependent 
anomalous subdiffusion. 

\section{Anomalous sub-diffusion Modeling}

A way to describe sub-diffusion is to start from a two dimensional random 
diffusion process.
A particle walks from trap to trap
and spend a certain (random) time in each trap. It is characterized by the 
following operation~:
\begin{equation}
{\bf r}\rightarrow{\bf r} + {\bf \Delta}; t \rightarrow t+\tau
\end {equation}
$\bf r$ and $t$ are respectively the two dimensional position and the age of 
the particle, where ${\bf \Delta }$
is a two dimensional random (Gaussian) variable with variance $v=2D$, and 
$\tau$ is the (random) time
the particle spend in the trap. 

In our model, the particle is supposed to diffuse very rapidly between two 
traps.
This travel time is therefore neglected. The time $\tau$ the particle stays 
in a trap is supposed to have very strong fluctuations, this give rise to 
anomalous diffusion pattern.

As an example a generic distribution is used which leads, after a while, to a 
standard Levy law in time~:
\begin{equation}
P_0(\tau)\ =\ {\alpha\over (1+\tau)^{\alpha+1}}
\end {equation}
This distribution have been used in the same type of context by Naggle 
\cite{b.j}.

The Levy exponent $\alpha$ is the characteristic exponent of subdiffusive 
behaviour. For long times we have : 
\begin{equation}
<r^2(t)> \propto t^{\alpha}
\end {equation}

When $\alpha < 1$ a spatio-temporal Fourier (Laplace) analysis leads to the 
following asymptotic ($\omega$, $k$ $\to$ 0) Green function~:
\begin{equation}
\tilde{g}({\bf k},\omega)\ =\ {1 \over \omega(D_\alpha k^2  
\omega^{-\alpha}+1)}\ ;\ D_\alpha={D/ \Gamma(1-\alpha)}
\label{l.f.}
\end {equation}
where $\omega$ and ${\bf k}$ are respectively the conjugate variables of 
position ${\bf r}$ and time $t$, where $k=|{\bf k}|$.
Notice that the solution of the inverse Laplace transform is a function of the 
variable $k^2t^\alpha$. It follows that the Green function is a function of 
the variable $x=r^2/t^\alpha$.
When $x$ is high enough one can perform an approximate inverse transformation 
via a saddle point method~:
\begin{equation}
g({\bf r},t) \  \propto \  \exp (-cst \ x^\nu) \  \  ;\   \nu={1\over 
2-\alpha},\  cst\  {\rm :\  a\  known\  constant}
\end {equation}
Notice that the exponent $\nu$ interpolate nicely between the gaussian case 
($\alpha=1$) and the exponential case.
The general solution of this type of anomalous diffusion process is then~:
\begin{equation}
\rho({\bf r},t)=\int \rho_0({\bf r'}-{\bf r}) g(x(r',t))\ d^2{\bf r'}
\end {equation}
where $\rho$ is the probability density to find the particle at the point 
${\bf r}$ at instant $t$ and $\rho_0$ is the initial state.

As the green function is a bell-shaped fast decreasing function, one  
approximate it by a gaussian shape with the exact dispersion, 
$\tilde{D}_\alpha=D \sin(\pi \alpha)/(\pi \alpha)$, which can be calculated 
from eq.\ref{l.f.}.  This permits to construct an analytical expression of the fluorescence recovery 
using standard properties of Gaussian functions. 

Starting from Axelrod \cite{b.f} initial density as it is immediately after a 
Gaussian laser beam profile extinction indeed :
\begin{equation}
\rho_0({\bf r})=\exp(-K \exp(-2{{\bf r}^2 \over R^2}))
\end {equation}
($K$={\rm photobleaching constant, depending on experimental conditions 
\cite{b.f}}) and using the standard properties of the Gaussian shape 
in the convolution operation, one can obtain the time evolved result as a serie.

Once integrated upon a disk of radius $R$, and,
after normalization to the surface of the disk, one obtain the FRAP signal~:

\begin{equation}
\label{recovery}
I_R(t)\ =\ 1+\sum_1^\infty {(-K)^n \over n!}{1\over 2 n}\left(1-\exp\left(-{2 
n R^2 \over R^2+4 n \tilde{D}_\alpha t^\alpha}\right)\right)
\end {equation}

This function will be used to fit experimental data.\\

Systematic corrections of this procedure are determined
using numerical Monte-Carlo simulations of the fluorescence recoveries, using known 
$\alpha$ and $\tilde{D}_\alpha$

A more precise study of the Green function will be published later.

\section {Experiments}

Experiments were conducted on Baby hamster kidney cells (BHK) grown on a 
coverslip in cell culture medium for 2 days. Cells were transfected 24 hours 
before the FRAP experiment with a pC1EGFPPHEFA6 plasmid. This plasmid contains 
the sequence for both PH-EFA6 domain and EGFP as a fluorescent label, linked 
to the N-terminus of the PH-EFA6 domain in order to avoid any perturbation to 
the membrane linkage. 
FRAP measurements were made with a commercially available confocal microscope. 
Prebleached images were firstly acquired to ensure for the lack of 
photo-destruction during the observation. A brief laser pulse was then 
delivered to the cell (500ms). Images were thereafter recorded at given 
intervals (440ms). The intensity ratio between the extinction laser beam and 
the monitoring laser beam was fixed to $10^6$. Each fluorescence recovery was 
recorded for 80 s at 25 C, containing 150 experimental values. Focus of the 
laser by a 63x objective produced a Gaussian intensity distibution of the beam 
in the object plane.

\begin{figure}
\epsfig{file=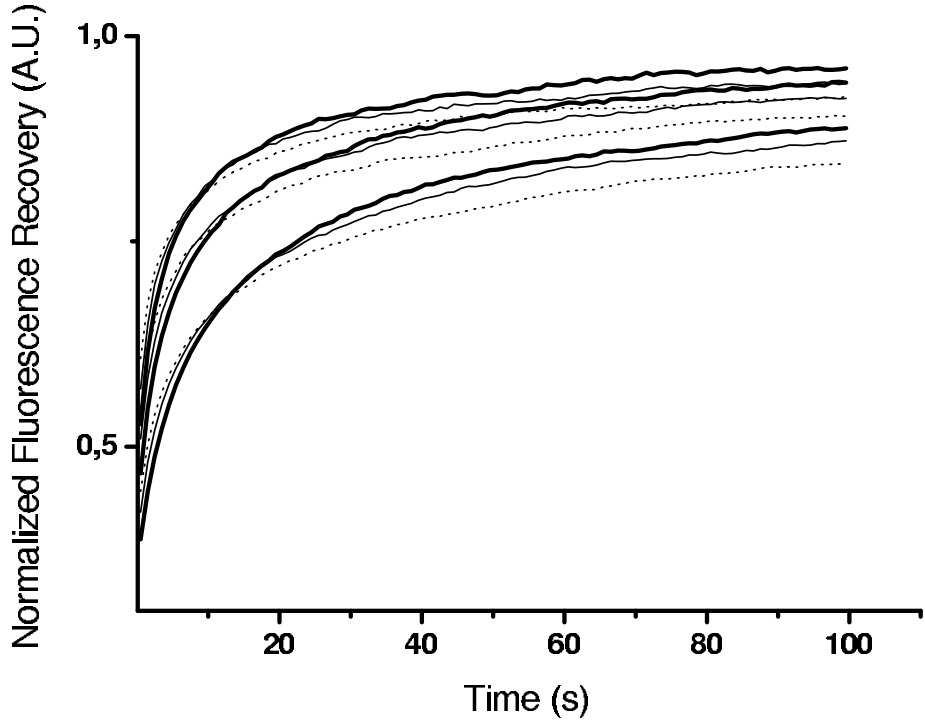, width=\textwidth}
\caption{{\bf Monte-Carlo simulation of normalized fluorescence recoveries in the cas of anomalous subdiffusion \bf} . {\it Different values of $D$ and $\alpha$
have been tested in the simulations. Here, values of $D = 0.5 ; 1 ; 1.5 $ are represented from bottom to top with different $\alpha$ in each case : 0.6 (dots) ; 
0.7 (thin line) ; 0.8 (thick line). The Monte Carlo has been constructed with $10^7$ individual trajectories \it}}
\label{f.1}
\end{figure}

\begin{table}
\begin{center}
\begin{tabular}{ccccccccc}
& $\alpha$ = 0.6&& $\alpha$  = 0.7&& $\alpha$  = 0.8 \\
\\
Input D&Fitted D&Correction&Fitted D&Correction&Fitted D&Correction \\
0.5&0.305&1.639&0.322&1.551&0.378&1.322\\
1&0.460&2.172&0.534&1.871&0.627&1.596\\
1.5&0.666&2.253&0.726&2.066&0.872&1.721\\
&&&&&&&\\
\end{tabular}
\caption[smallcaption] {{ \bf Correction factors for D.\bf} Correction factors have been obtained by fitting Monte-Carlo simulations of normalized 
fluorescence recoveries with a $10^th$ order limited development of the analytical expression established here (eq.\ref{recovery})}. 
\label{t.1}
\end{center}
\end{table}

\section {Results}

In order to validate our model, Monte-Carlo simulation of the fluorescence 
recovery have been made using different values of $\alpha$ and different value 
of $D$ (see fig. \ref {f.1}) with K=4 in every case since this value of K was the one 
found in FRAP experiments . 
These simulations have then been 
fitted using a $10^{th}$ order limited development of the fluorescence 
recovery equation established for anomalous diffusion (eq. \ref {recovery}).
Input $\alpha$ of respectively 0.6, 0.7, 0.8 led to fitted value of $0.5 \pm 
0.02$, $0.65 \pm 0.03$ and $0.75 \pm 0.03$. Absolute error measured on 
$\alpha$ is found between $9 \pm 3 \%$ and $6 \pm 3\%$, decreasing with
increasing $\alpha$.
Values found for $D$ are further from those input in our simulations (ranging 
from 0.25 to 10). For this reason, results were calibrated using appropriate 
factors for the three values of $\alpha$. A comparison of input and found 
values of $D$ as well as some correction factors are given in Table \ref {t.1}. 

\begin{figure}
\epsfig{file=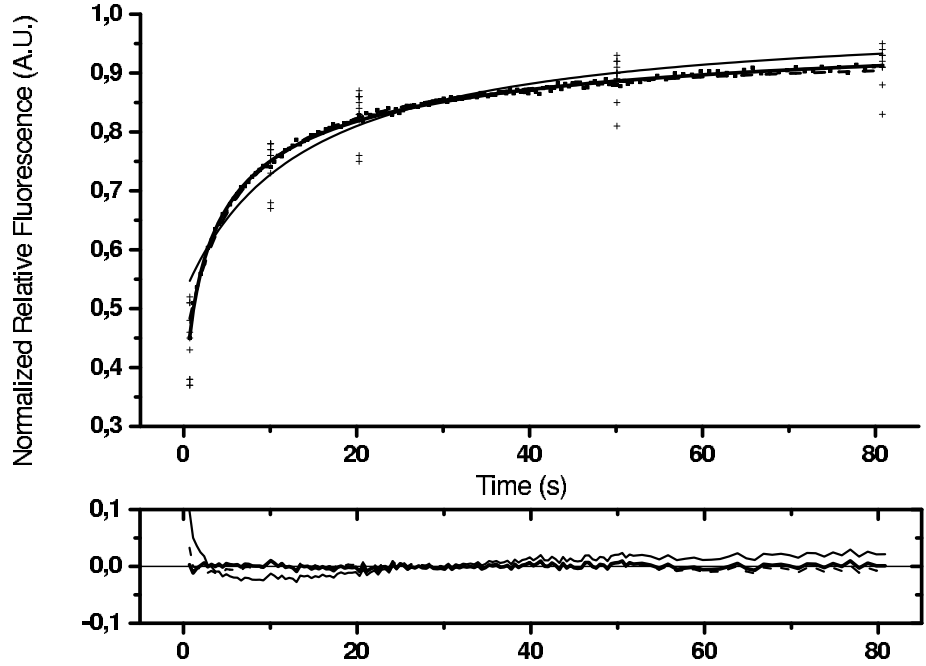, width=\textwidth}
\caption{\bf Fit of the experimental normalized fluorescence recoveries : \bf
\it Upper part : Experimental curve fitted using the three different models 
(see text for explanation).  Lower part : Difference between experimental values and the three models : $Fc - Fo$. Thin line : Normal diffusion without immobile fraction, Dashed line : Normal diffusion with immobile fraction, Thick line : Anomalous subdiffusion.\it} 
\label{f.2}
\end{figure}

Experimental fluorescence recoveries of the protein obtained in 
living cells (n=17) were fitted by three different models (Fig.\ref {f.2}):
\\ - classical diffusion without restriction (derived from eq.12 of Axelrod 
\cite{b.f}).
\\ - classical diffusion introducing an immobile fraction M. For this model 
eq.12 from Axelrod \cite{b.f} has been derived, leading to the 
following expression of $I_R(t)$ : 

\begin{equation}
\label{axelmob}
I_R(t)\ =\ {1-\exp(-K) \over K}(1-M)+M \sum_1^\infty{(-K)^n \over n!(1+n+8 n D 
t / R^2)}
\end{equation}
\\ - time dependent anomalous diffusion (for which the analytical fluorescence 
recovery has been established in the modeling section (eq.\ref{recovery})).\\
The quality of the fit was estimated using a $\chi^2$ statistical test.

\begin{table}
\begin{center}
\begin{tabular}{cccccc}
$\alpha$&$M$&$D(~\mu m^2.s^{-1})$&$\tilde{D}_\alpha (~\mu m^2.s^{-\alpha})$&$\chi^2$\\
-&1&$0.119 \pm 0.061$&-&$5.7 \pm 0.5$ \\
-&$0.917 \pm 0.004$&$0.217 \pm 0.010$&-&$3.8 \pm 1.6$ \\
$0.63 \pm 0.02$&-&-&$1.48 \pm 0.05$&$2.9 \pm 1.6$ \\
\\

\end{tabular}
\caption[smallcaption] { {\bf Experimental values obtained with the different models. \bf} \it Values obtained after fit of the experimental recovery 
using the three different models described in the text. \it}
\label{t.2}
\end{center}
\end{table}

Table \ref{t.2} shows that fitting the experimental curves with the classical model of Axelrod 
led to very bad results, whereas using a limited development to the $10^{th}$ order of  
eq.(\ref {recovery}) and eq.(\ref{axelmob}) led to slightly different quality of the fit.

Once corrected by the previously determined factor, the calculated value of D for anomalous
sub-diffusion using the expression $D=\tilde{D}_\alpha (\pi \alpha)/ \sin(\pi \alpha)$ gives : 
$D = 10.4 \pm 0.7 ~\mu m^2.s^{-\alpha}$.

It has to be noted that value of D found using the two models are (intrinsically) different. 
However they can be compared according the relation : $D(t)=Dt^{\alpha-1}$ \cite {b.l, b.m}.
Therefore, if one estimate D(t) at time of half recovery of fluorescence ($t=14,4s$), as it is usually done for the estimation of D using classical diffusion model 
\cite {b.f, b.g, b.l} a value of $D(14,4) =  3.9 \pm 0.4~\mu m^2.s^{-1}$ can be found which is more than ten time higher than the
 value found using classical diffusion with an immobile fraction, $D = 0.217 \pm 0.010~\mu m^2.s^{-1}$.

\section {Discussion}

In this paper, anomalous subdiffusion in fluorescence recoveries experiments
have been reexamined from the beginning. An analytical formulation of the recovery 
curve as a function of time have been calculated using a Gaussian extinction profile.
This equation has been tested by fitting Monte-Carlo simulated fluorescence recoveries. 
As already observed by Feder {\it et al.\it} \cite {b.k}, $\alpha$ was underestimated in the fit.
This could be explain by asymptotic effects occuring because of the time-scale of the experiment \cite{b.j}.
More surprinsingly, $\tilde{D}_\alpha$ was also itself underestimated. 
This has not been underlined by Feder {\it et al.\it} \cite {b.k}, but seems to be crucial for correct estimation of $D$.
Combination of these two parameters led us to established correction factors, 
depending both on $\alpha$ and $\tilde{D}_\alpha$ to achieved measurement of $D$.

The experimental recoveries obtained on PH-EFA6 showed that Brownian diffusion 
without immobile fraction did not fit the data. This strongly suggest that 
this membrane bound protein is also submitted to restricted motion at the 
surface of the membrane. Therefore, it was interesting to analyze the 
recoveries using the two limit models of restricted motion. Whereas this work show that it is not possible to firmly distinguish between free-diffusion with an 
immobile fraction and anomalous diffusion of the entire set of proteins only by statistical considerations on the quality of the fit,
estimation of $D$ at half time recovery obtained using anomalous diffusion ($D \simeq 4~\mu m^2.s^{-1}$) 
led to a value close to that obtained for lipids in fluid-phase model membranes (lipid bilayers without proteins). 
This value would mean that diffusion of this protein from trap to trap is mainly due to its lipid links. 
This hypothesis is acceptable regarding biochemical data obtained on interaction of this protein domain with lipid membrane.

While analysis of SPT measurements using anomalous diffusion is now becoming usual  
it is still difficult to elucidate its relevance in FRAP experiments. Indeed, as a limit of the technique, 
FRAP inherently averages over a large number of particles. This could explain why anomalous diffusion model 
(in which complete but restricted diffusion is allowed) and classical Brownian motion with an 
immobile fraction (in which free diffusion occurs for one subpopulation and no 
diffusion for the other subpopulation) lead to the same statistical quality 
for fitting experimental data. Therefore there is still a challenge in trying to agreement SPT 
data with FRAP data on biological molecules {\it  in situ \it}.

\section {Acknowledgments}
Authors would like to acknowledge M. Franco for kindly giving us the plasmid 
used for PH-EFA6 expression in cell lines and M. Partisani for cell culture and 
transfection. They are also indebted to A. Lopez for critical reading of the manuscript
and fruitfull discussions.

\end{document}